\documentclass{Interspeech2024}
\usepackage{graphicx}
\usepackage{subcaption}
\usepackage{float}
\usepackage{cite}
\usepackage{multicol}
\usepackage{amsmath}
\usepackage[table]{xcolor}
\usepackage{booktabs}
\usepackage{amssymb} 
\usepackage{caption}
\usepackage{pifont}
\interspeechcameraready

\title{Binaural Selective Attention Model for Target Speaker Extraction}

\name[affiliation={1}]{Hanyu}{Meng}
\name[affiliation={1}]{Qiquan}{Zhang}
\name[affiliation={1}]{Xiangyu}{Zhang}
\name[affiliation={1}]{Vidhyasaharan}{Sethu}
\name[affiliation={1}]{Eliathamby}{Ambikairajah}

\address{
  $^1$The University of New South Wales, Australia}
\email{\{hanyu.meng, qiquan.zhang, xiangyu.zhang2, v.sethu,  e.ambikairajah\}@unsw.edu.au}

\keywords{binaural modelling, cocktail party problem, speaker extraction, selective attention}

\begin{document}

\maketitle

\begin{abstract}
The remarkable ability of humans to selectively focus on a target speaker in cocktail party scenarios is facilitated by binaural audio processing. In this paper, we present a binaural time-domain Target Speaker Extraction model based on the Filter-and-Sum Network (FaSNet).  Inspired by human selective hearing, our proposed model introduces target speaker embedding into separators using a multi-head attention-based selective attention block. We also compared two binaural interaction approaches -- the cosine similarity of time-domain signals and inter-channel correlation in learned spectral representations. Our experimental results show that our proposed model outperforms monaural configurations and state-of-the-art multi-channel target speaker extraction models, achieving best-in-class performance with 18.52 dB SI-SDR, 19.12 dB SDR, and 3.05 PESQ scores under anechoic two-speaker test configurations.

\end{abstract}
\section{Introduction}
\begin{figure}[t]
\centering
    \begin{subfigure}[b]{\columnwidth}
        \centering
        \includegraphics[width=0.8\textwidth]{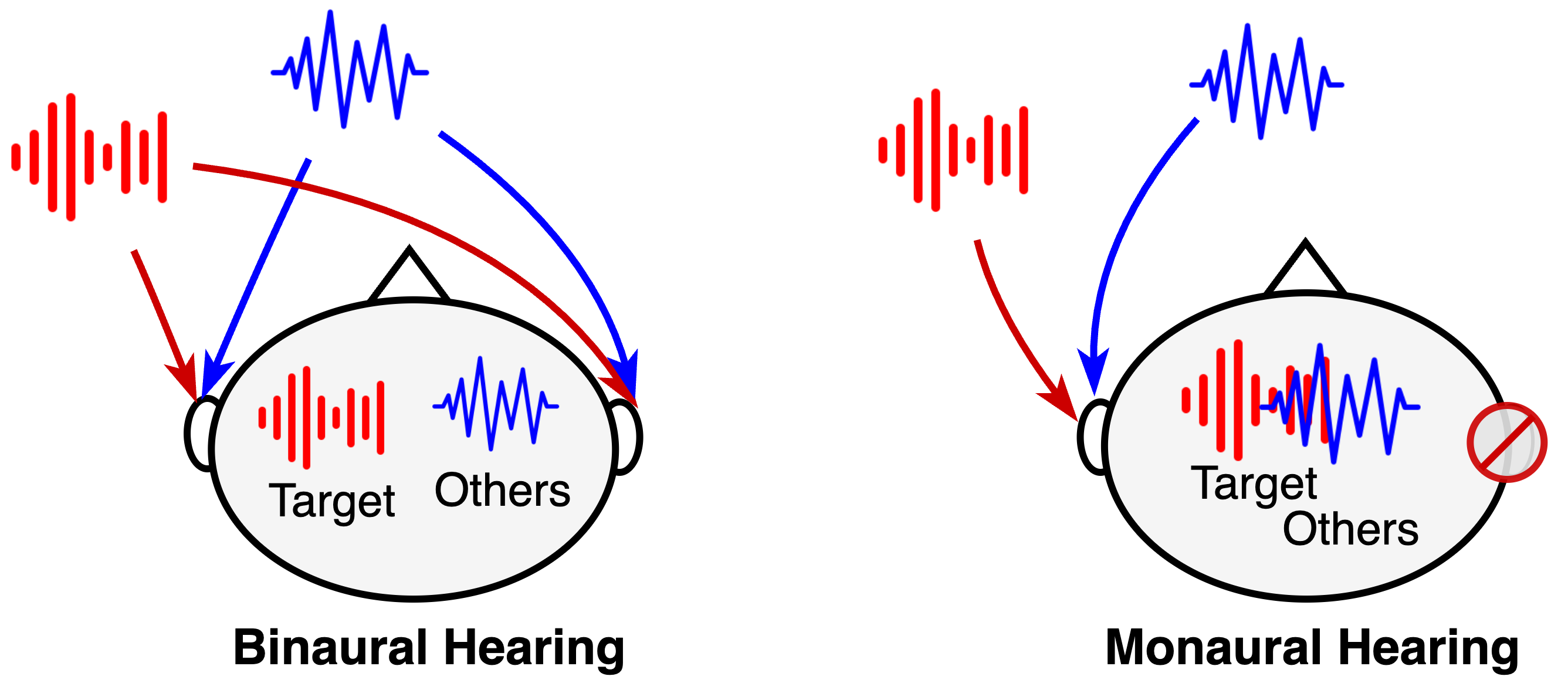}
        \caption{The significance of binaural hearing in separate the target source}
        \label{fig:sub1}
    \end{subfigure}
    \hfill 
    \begin{subfigure}[b]{\columnwidth}
        \centering
        \includegraphics[width=0.75\textwidth]{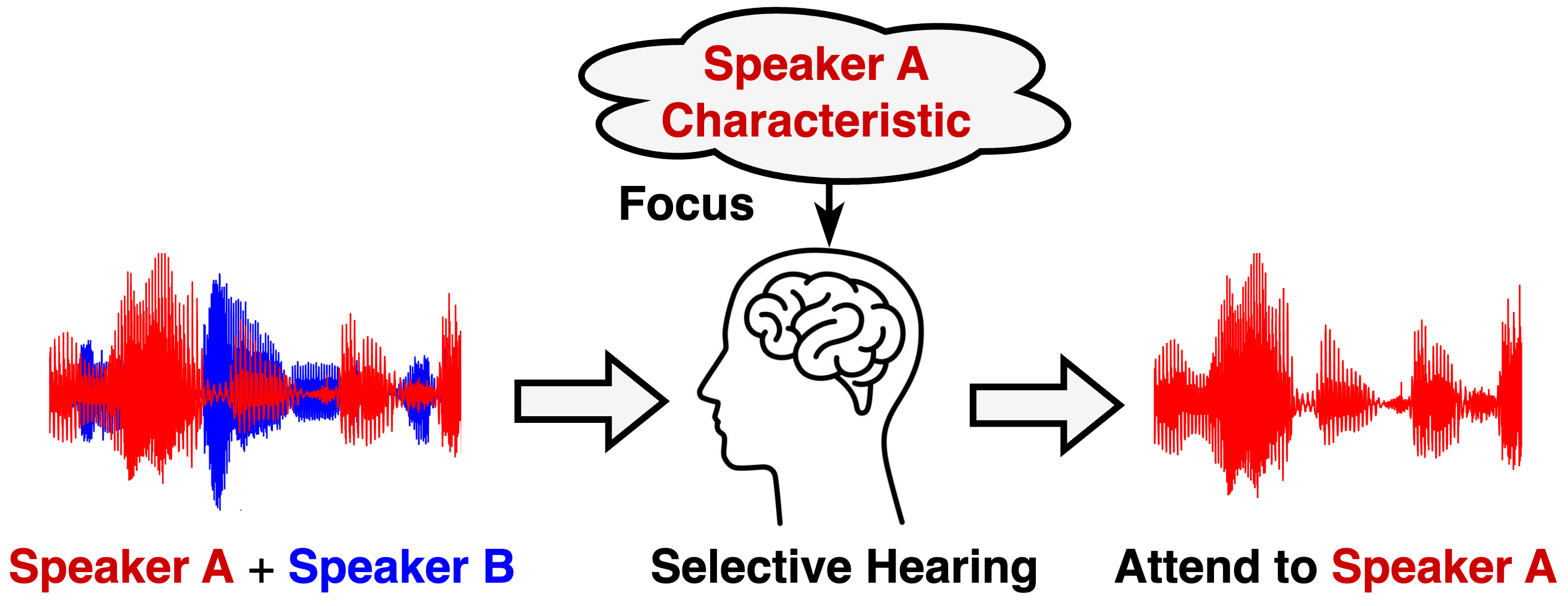}
        \caption{The selective hearing through attention and the prior knowledge of the target source}
        \label{fig:sub2}
    \end{subfigure}
    \caption{An illustration of binaural selective attention}
    \label{fig:binaural_selective_att}
    \vspace{-0.5cm}
\end{figure}

The `cocktail party problem' describes the human capability for binaural auditory selective attention to focus on a target speaker in an environment containing several speakers and other ambient noise~\cite{cocktail}. This phenomenon, grounded in psychoacoustic research, reveals that such auditory discernment arises from the interplay of selective hearing mechanisms and binaural auditory processing~\cite{cocktail2}. As shown in Figure~\ref{fig:sub1}, binaural hearing exceeds monaural hearing by picking up spatial cues from sound sources, which helps us identify and focus on a specific speaker in an environment containing various sounds. Furthermore, Figure~\ref{fig:sub2} illustrates that selective attention is determined by prior knowledge of the target speaker. This process guides our 
`top-down' attention, allows us to concentrate on the voice of the person we want to listen to and ignore other unnecessary sources~\cite{auditory_att}.

In computational auditory scene analysis (CASA), numerous efforts have sought to emulate this selective hearing capability~\cite{casa}. Early research often frames CASA as a single-channel speech separation task~\cite{monaural_separation,complex_ratio_masking,deepmmse,tfaj}, ignoring the nature of two ear inputs for human auditory processing. While several recent studies have considered the speech separation in microphone array scenarios~\cite{wu2020end,9747261,zhang2021multi}, they often fail to replicate the authentic spatial configuration of human binaural hearing and do not adequately model the shadowing effects contributed by the head and pinna.

Recent advancements have narrowed the broad challenge of selective hearing modelling to a more precise task known as Target Speaker Extraction (TSE)\cite{tse_overview}. TSE simulates the human ability to focus on a particular speaker by using clues about who to listen to. Advanced TSE systems like SpeakerBeam\cite{speakerbeam1}, SpEx~\cite{spex}, and VoiceFilter~\cite{voicefilter} start by analysing an enrollment speech of the target speaker. They create unique speaker embeddings from this enrollment and then use these patterns to guide a network that can separate the target voice from the mixture input. The output of TSE systems is an estimated stream of the target speaker's content in the mixture, while the speech of others is turned down or removed~\cite{add_spk_embedding,speaker_representation}.

While some TSE models use spatial clues like the Spatial Neural Filter model~\cite{neural_spatial_filter}, and visual clues, particularly useful in teleconferencing and in-vehicle systems~\cite{ochiai2019multimodal}, human auditory perception often relies on speaker identity, enabled through binaural interactions for spatial localisation. Crucial auditory cues, including Interaural Time Difference (ITD), Interaural Intensity Difference (IID), and Interaural Correlation (IC), play a significant role in spatial orientation and speaker separation by utilising the differences captured by both ears~\cite{hawley1999speech,itd}.

To capitalise on inter-channel information for speech extraction, beamforming techniques have been extensively applied to separate target sources in a specific direction. Those methods are based on a filter-and-sum process to distinctly isolate individual audio sources. A significant body of research has incorporated beamforming as a foundational framework, integrating deep neural networks (DNNs) to optimise filter weights~\cite{learnable_beamforming,wang2021sequential}. This integration addresses the limitations of spatial resolution associated with a limited number of microphones, thereby enhancing separation performance. Therefore, our study applies a time-domain neural beamforming based system as the separator in our proposed model. This choice is predicated on the method's proven efficacy in leveraging temporal dynamics and spatial cues for more accurate source separation~\cite{tac_fasnet}.

This paper explores a binaural selective attention model for target speaker extraction, utilising the Filter-and-Sum network (FaSNet)~\cite{fasnet} as a foundation and simulating ear inputs with Head-Related Transfer Functions (HRTFs). We design a selective attention block for adapting speaker embedding to the separator. We also explore binaural interaction methods, focusing on cosine similarity (CSim) for time domain signal and inter-channel attention correlation (IAC) in learned spectral representation. This leads to two binaural target speaker extraction models, which we refer to as Bi-CSim-TSE and Bi-IAC-TSE, which are both presented in the following section.

\section{System Overview}
\subsection{Problem Formulation}
\subsubsection{Binaural Received Signal Model}
Given \(C\) sources in space, denoted as \(\textbf{s}_i\), where \(i \in \{1,2,\ldots,C\}\), the binaural received signals at the left and right ears can be expressed in a unified manner as:
\begin{equation}
\setlength{\abovedisplayskip}{3pt}
\setlength{\belowdisplayskip}{3pt}
\textbf{x}_{\{L,R\}} = \sum_{i=1}^{C} \textbf{s}_i * a^{\{L,R\}}_{i} + \mathcal{N}_{\{L,R\}}
\end{equation}
where \(\textbf{s}_i = [s_i[0], s_i[1], \ldots, s_i[T-1]]\) represents the signal from the \(i\)-th source with a length of $T$, and \(\mathcal{N}_L\) and \(\mathcal{N}_R\) denote the diffuse noise at the left and right ears, respectively. The signals are convolved with the HRTFs corresponding to the direction of the source. Assuming the sources are at the same elevation, the direction corresponds to the azimuth on a 2D plane. The HRTF for a given direction \(\theta\) can be defined as:
\begin{equation}
\setlength{\abovedisplayskip}{3pt}
\setlength{\belowdisplayskip}{3pt}
A_{\theta} = \{a_{\theta}^{L}, a_{\theta}^{R}\}
\end{equation}
where \(\theta \in \{-90^\circ, -90^\circ + \Delta\theta, \ldots, 90^\circ - \Delta\theta, 90^\circ\}\), \(\Delta\theta\) represents the resolution of discretization in degrees. Here, \(a_{\theta}^{L}\) and \(a_{\theta}^{R}\) represent impulse responses of length \(L\).
\subsubsection{Aims}
The objective of the binaural selective attention model is to construct a system capable of processing the received signals at both ears, along with a priori knowledge of the target source characteristics \( c_{\text{target}} \), to estimate the system \( \mathcal{H} \) with binaural parameters \( \theta_{\text{binaural}} \). This estimation can be formalised as:
\begin{equation}
\textstyle
\hat{\textbf{x}}_{\text{target}} = \mathcal{H}(\textbf{x}_L, \textbf{x}_R, c_{\text{target}}; \theta_{\text{binaural}})
\end{equation}
\subsection{Feature Extraction}
\subsubsection{Spectral Feature}
\begin{figure*}[ht]
\centering
\includegraphics[width=0.9\textwidth]{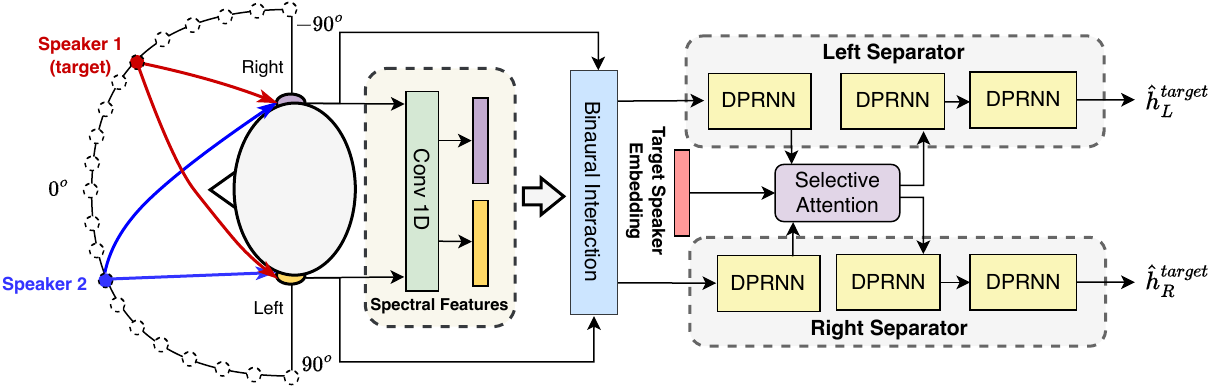}
\caption{An Overview of the proposed binaural target speaker extraction model}
\label{sys_overview}
\end{figure*}

In this study, we focus on the time domain TSE approach, exploiting a one-dimensional (1D) convolutional layer designed to approximate the linear transformation characteristic of the input speech signal as illustrated in Figure~\ref{sys_overview}.

Given that we have a frame of signal with length $m$, $\bold{x}_{k,i}[n]$, where $n\in\{1,2,..,m\}$, and $i\in\{{1,...,N}$\}, $N$ is the total frame numbers. The spectral feature can be expressed as
\begin{equation}
 C_{k,i} = \text{Conv1D}\;(\bold{x}_{k,i}), \quad k = 1,2
\end{equation}
Here we choose the left ear as the reference channel, corresponding to the case $k=1$, and $k=2$ representing the right ear.
\subsubsection{Binaural Interaction}

For the reference ear, indicated by $k=1$, we compute the Cosine Similarity (CSim) by comparing the central frame $\bold{x}_{1,i}$ with its subsequent segment $\bold{d}_{1,i}$. This comparison starts at the initial index of the segment. The process involves a stepwise progression, where the segment is shifted one sample at a time until the end of the segment is reached. This iterative process can be expressed as
\begin{equation}
\text{CSim}_{k,i}[j] = \frac{\bold{x}_{1,i} \cdot (\bold{d}_{k,i}[(j-1)N+1:jN])^T}{||\bold{x}_{1,i}|| \cdot ||\bold{d}_{k,i}[(j-1)N+1:jN]||}
\end{equation}
where $k=1,2$ represents the left and right ears respectively, and $j = 1, \ldots, K-N+1$ to account for each possible alignment within the segment relative to the central frame.
The reference ear's CSim measures self-similarity between a frame and its context. Such self-similarity is crucial for distinguishing voiced speech from silence. Voiced speech has distinct, repetitive waveforms easily detected by CSim, while silence shows low similarity, indicating minimal informational content to the separator. As for the right ear, the CSim feature gives binaural similarity between the selected left frame $\bold{x}_{1,k}$ and the corresponding right segment $\bold{d}_{2,i}$. We named the model
with this CSim binaural feature as the Bi-CSim-TSE model. It is worth noting that our CSim feature differs from FaSNet's normalised cross-correlation (NCC)~\cite{fasnet} by parallel calculating the CSim of left and right channels, instead of first getting the pre-separation output of the reference ear. 

The second interaction approach is adopted from~\cite{iac} named inter-channel attention correlation (IAC). Different from CSim, which derives directly from the time domain binaural signal, IAC uses the learnable spectral feature and interacts with the matrix multiplication followed by a softmax layer, which can be given by 
\begin{figure}[ht]
\centering
\includegraphics[width=\columnwidth]{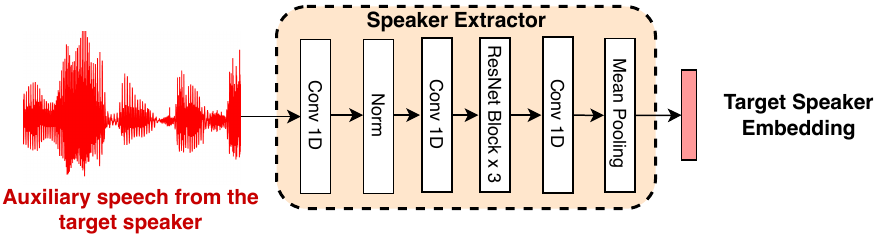} 
\caption{The structure of the speaker extractor to get the target speaker embedding}
\label{spk_extractor}
\vspace{-0.5cm}
\end{figure}
\begin{equation}
\setlength{\abovedisplayskip}{3pt}
\setlength{\belowdisplayskip}{3pt}
\text{IAC}_{i}[j] = \text{Softmax}(C_{1,i}C_{2,i}^{T})
\end{equation}
In this approach, both left and right binaural features are identical. Unlike CSim's direct time-domain analysis, the IAC uses learned spectral features from the encoder. It employs matrix multiplication followed by a softmax layer, highlighting relevant features for identifying the target speaker's spatial location while diminishing less relevant signals. We named the model with this binaural interaction way as the Bi-IAC-TSE model.
\subsection{Speech Separator}
As shown in Figure~\ref{sys_overview}, the separators receive their inputs from a combination of spectral and binaural features for each auditory channel, as previously described. We employ three sequential blocks of the Dual-Path Recurrent Neural Network (DPRNN)~\cite{dprnn} for the separation tasks of both the left and right audio channels. The DPRNN is known for its effectiveness in capturing the temporal dynamics crucial for separating speech signals. Within the DPRNN framework, we utilise bidirectional Long Short-Term Memory (BLSTM) networks as the core recurrent neural network (RNN) components. We also apply the Transform-Average-Concatenate (TAC) strategy~\cite{tac_fasnet} between the first DPRNN block for left and right separator. 

The separators' ultimate output is characterised by impulse responses, specifically $\hat{\bold{h}}_L^{target}$ and $\hat{\bold{h}}_R^{target}$, which act as beamforming filters.
\subsection{Selective Attention with Speaker Embedding}
The speaker encoder's network structure, illustrated in Figure~\ref{spk_extractor}, effectively generates speaker embeddings for voice classification and has been utilised in various TSE models~\cite{spex,spex+,wang2021neural}. It converts speaker enrollment utterance into an embedding that captures the speaker's unique voice traits.

To integrate the speaker embedding within our model, we standardise the dimensions of various inputs to the selective attention block - specifically, the speaker embedding, the first block outputs from DPRNN, and the outputs from the TAC module for both separators. 

The model employs a selective attention mechanism, treating the speech embedding $Y_{i}$ as the key, DPRNN output $H_{i}$ as the queue, and TAC output $G_{i}$ as the value, leading to the construction of a multi-head self-attention (MHSA) block~\cite{transformer,10446337}:
\begin{equation}
\textstyle
O_{i} = \text{MHSA}(H_{i}, Y_{i}, G_{i})
\label{attention}
\end{equation}
The output, $O_{i}$, is then propagated as input into the subsequent DPRNN block for both separators.
\subsection{Reconstruction}
As indicated in Figure~\ref{sys_overview}, the extraction model generates left and right ear beamforming filters, using a filter-and-sum operation to estimate the target signal for frame $i$.
\begin{equation}
\textstyle
    \hat{\bold{y}}_{i,target} = (\hat{\bold{h}}_{L}^{target}\ast \bold{x}_{1,i}+\hat{\bold{h}}_{R}^{target}\ast \bold{x}_{2,i})/2
    \label{fas}
\end{equation}
For full signal reconstruction, the overlap-add technique is applied across all $m$ frames.
\subsection{Loss Function}
The end-to-end training utilises scale-invariant signal-to-distortion ratio (SI-SDR) separation loss, jointly optimising the speech and speaker extractors. The loss function is defined as follows
\begin{equation}
\mathcal{L}_{\text{SI-SDR}} = -20\log_{10} \frac{||(\hat{\bold{x}}^T\bold{x}/\bold{x}^T\bold{x})\cdot\bold{x}||}{||(\hat{\bold{x}}^T \bold{x}/\bold{x}^T \bold{x})\cdot\bold{x} - \hat{\bold{x}}||}
\end{equation}

where $\hat{\bold{x}}$ and $\bold{x}$ are the estimated signal and the target clean signal, respectively. Their means are normalised to zero.

\section{Experiments}

\begin{table*}[htbp]
\centering
\caption{Extracted speech SDR (dB), SI-SDR (dB), PESQ, and STOI for the proposed models and modified existing TSE models.}
\begin{tabular}{lcccccc}
\toprule
\textbf{Method} & \textbf{Speaker Embedding} & \textbf{Input Type} & \textbf{SI-SDR (dB)} & \textbf{SDR (dB)} & \textbf{PESQ} & \textbf{STOI} \\
\midrule
Mixture Left & - & Monaural & -13.51 & 0.6 & 1.17 &  0.5 \\
Mixture Right & - & Monaural & -14.41 & 0.6 & 1.2 & 0.5 \\
Modified SpEx+~\cite{spex+} & \checkmark & Monaural & 3.16 & 5.84 & 1.44 & 0.74\\
Monaural-TSE (Ours) & \checkmark & Monaural & 3.49 & 7.88 & 1.71 & 0.82\\
FasNet-TSE~\cite{tac_fasnet} & \checkmark & Binaural & 16.46 & 17.09 & 2.70 & 0.94\\
Modified TD-SpkBeam~\cite{td_speakerbeam} & \checkmark & Binaural & 17.48 & 18.20 & 2.98 & \textbf{0.96}\\
Bi-IAC-TSE (Ours) & \checkmark & Binaural & 17.38 & 18.06 & 2.93 &  0.95\\
Bi-CSim-TSE (Ours) & \checkmark & Binaural & \textbf{18.52} & \textbf{19.12} & \textbf{3.05} & \textbf{0.96}\\
\bottomrule
\end{tabular}
\label{results_table}
\vspace{-0.5cm}
\end{table*}
\subsection{Datasets}
In the generation of the binaural training dataset, single-channel speech signals are convolved with HRTFs to simulate spatial auditory scenes.  For this purpose, we utilise the Surrey HRTF dataset~\cite{hrtf}, which is measured using the Knowles Electronic Manikin for Acoustic Research (KEMAR) dummy head~\cite{kemar}. This dataset contains HRTFs in azimuth range from $-90^{\circ}$ to $90^{\circ}$, with a resolution of $5^{\circ}$. 

Concerning the speech corpus, we use the LibriSpeech dataset~\cite{librispeech}. In data pre-processing, all speech signals are truncated or extended to a duration of $4$ seconds with a $16$kHz sampling rate. The overlap ratio between dual speech signals is uniformly distributed from $0\%$ to $100\%$, ensuring an average overlap of $50\%$ across the dataset. Adjustments are made to align the two speech signals appropriately and normalise their volumes to achieve a random relative SNR ranging between 0 and 5 dB as outlined in~\cite{tac_fasnet}. The train, validation, and test set contains 20000, 5000, and 3000 samples correspondingly. No speakers overlap among all three sets. 

For the spatial attribute of the dataset, the azimuth of each speaker is randomly selected within the interval of $-90^{\circ}$ to $90^{\circ}$ to ensure a uniform distribution across the dataset. Auxiliary speech for the target speaker is randomly selected from Librispeech, excluding the reference utterance used in the speech mixture, and is aligned to $8$ seconds through either trimming or padding to conform to dataset specifications.
\begin{figure}[ht]
\centering
\includegraphics[width=\columnwidth]{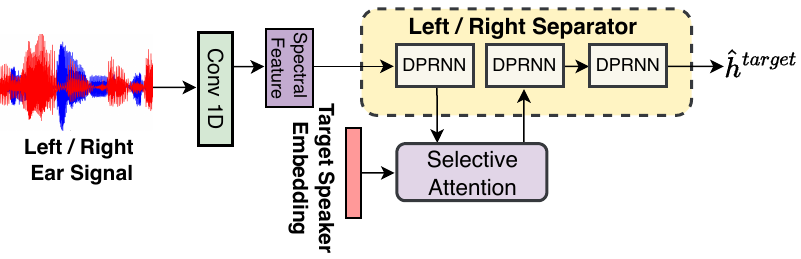}
\caption{The monaural configuration}
\label{monaural_tse}
\vspace{-0.3cm}
\end{figure}
\subsection{Model Configurations}
In our study, we conducted experiments on two binaural models with different binaural features: Bi-Sim-TSE and Bi-IAC-TSE.
To highlight the significance of binaural hearing over monaural hearing, we also made a monaural version of the proposed model as depicted in Figure~\ref{monaural_tse}. This configuration is different to the single-channel Target Speaker Extraction (TSE) system, where the monaural model processes either the left or right channel signal during the training and evaluation phases. This version of the model incorporates a single separator, receiving solely the spectral feature as its input. As for the selective attention block, TAC would not be applied, instead, the value of the multi-head attention is the same as the queue. 

To evaluate our proposed model, we compared it with the modified version of state-of-the-arts, including SpEx+~\cite{spex+} (single channel TSE), FasNet~\cite{fasnet}, and TD-Spkbeam~\cite{td_speakerbeam} for multi-channel TSE system. Here, we adapt FaSNet for TSE task, referred to as FaSNet-TSE.

For SpEx and TD-SpkBeam, we get rid of the multi-task learning scheme together optimised with a speaker classification system, but we reserve the speaker embedding extraction part together optimised with the separation network.
This modification was necessary because there's no speaker overlapping between train and validation in our experiment dataset, which is different from the WSJ0-2 mix dataset those work used originally. The dataset difference makes it challenging to train a speaker classification network.

As for the feature extraction part, we choose the segment size $K$ as $36$ms (i.e. $576$ samples at $16$Hz sampling rate). The frame size $N$ is $4$ms ($64$ samples). We set the speaker embedding and spectral feature dimension from the speaker extractor are the same, which is aligned as $64$. As for the multi-head attention for feeding the speaker embedding, we applied the 6-head attention block. For the extractor, our proposed model adheres to the same model parameter specifications as FasNet~\cite{fasnet,tac_fasnet}. 
\vspace{-0.2cm}
\section{Results and Discussion}
Table~\ref{results_table} presents the results for both binaural and monaural configurations against the modified version of existing models using widely accepted metrics in TSE tasks such as Scale-Invariant Signal-to-Distortion Ratio (SI-SDR), Signal-to-Distortion Ratio (SDR), Perceptual Evaluation of Speech Quality (PESQ), and Short-Time Objective Intelligibility (STOI). These metrics, along with tests on simulated left and right ear mixes, serve as benchmarks.

Our models, Monaural TSE, Bi-IAC-TSE, and Bi-CSim-TSE, incorporate a selective attention block for integrating speaker embeddings, demonstrating superior performance over the modified SpEx+ and TD-SpkBeam models, which removed the multi-task learning scheme. This underscores the effectiveness of our selective attention mechanism in utilising speaker embeddings to guide target speaker extraction. Contrasting with the state-of-art multi-channel speaker extraction model TD-SpkBeam, the most apparent difference between TD-SpkBeam and our model is that TD-SpkBeam estimates the mask for the target source and applies masking on the latent representation of the mixture input. However, our model aims to estimate beamformers to do time-domain binaural beamforming. We can conclude from the result that the time domain beamforming separator architecture gives better extraction performance as the estimated filter is directly applied to the time domain signal by filter-and-sum, avoiding unexpected error accumulation in the neural network during decoder reconstruction and masking.

Binaural configurations outperform monaural setups, highlighting the importance of dual-channel information and binaural interaction, consistent with psychoacoustic studies that suggest binaural hearing provides a more than $7$dB significant advantage over monaural hearing in one cocktail party environment~\cite{hawley2004benefit}. Furthermore, the comparison of our models to FasNet-TSE with two microphone settings shows the benefits of our proposed binaural interactive approach upon NCC based channel interactive ways depending on the pre-separation waveform in FasNet.

Among the proposed binaural models, Bi-CSim-TSE outperforms Bi-IAC-TSE, confirming that cosine similarity captures frame characteristics and location more effectively than IAC. It can explained that time-domain analysis offers clearer preservation of the original spatial and temporal characteristics of the signal than learned latent representation from the speech encoder. This can help with the following time domain beamformer extractor to learn the repetitive and binaural differences, therefore inference the target source's spatial information.

\vspace{-0.2cm}
\section{Conclusion}
In this paper, we modelled the binaural selective hearing process as a target speaker extraction task based on FaSNet. The experimental results show that the binaural model is superior to the monaural model in terms of speech extraction results. Furthermore, the reference ear and binaural cosine similarity features are shown to be a more effective binaural interaction feature than concatenating the encoded spectral feature through softmax. We also proposed a novel and effective way to feed the speaker embedded into the speaker extractor network through a multi-head attention guide for the speaker extraction from the mixture speech without multi-task learning. Experiments show that our proposed Bi-CSim-TSE achieved significant performance improvement, which proved to be an effective way to model the binaural selective hearing process. Finally, the comparison between cosine similarity based binaural interaction and interchannel attention correlation based binaural interaction revealed that the cosine similarity based approach was superior. 

\vspace{-0.2cm}
\section{Acknowledgements}
This work was funded by the Australian Research Council Discovery Grant DP210101228.


\bibliographystyle{IEEEtran}
\bibliography{mybib}

\end{document}